\documentclass{article}
\usepackage{amsmath,amssymb,amsthm}
\usepackage[usenames]{color}
\usepackage{url}
\usepackage{hyperref}
\newcommand{\CloseRelativeOfJohn}{\mathrm{CloseRelativeOfJohn}}
\newcommand{\CloseFriendOfJohn}{\mathrm{CloseFriendOfJohn}}
\newcommand{\F}{\mathcal{F}}
\newcommand{\In}{\mathrm{In}}
\newcommand{\Owns}{\mathrm{Owns}}
\renewcommand{\P}{\mathtt{P}}
\renewcommand{\phi}{\varphi}

\newtheorem{proviso}{Proviso}

\theoremstyle{definition}
\newtheorem{example}{Example}

\newcommand{\A}{\smallskip\noindent\textbf{A:\ }}
\newcommand{\Q}{\smallskip\noindent\textbf{Q:\ }}

\begin{document}
\title{Impugning Randomness, Convincingly}
\author{Yuri Gurevich%
\thanks{Microsoft Research, Redmond, WA, U.S.A. \url{gurevich@microsoft.com}}
\and
Grant Olney Passmore%
\thanks{Clare Hall, University of Cambridge and LFCS, University of Edinburgh, UK.
\url{grant.passmore@cl.cam.ac.uk}}
\date{May 2011}}
\maketitle

\mbox{}\vspace{1in}
\begin{abstract}\noindent
John organized a state lottery and his wife won the main prize. You may feel that the event of her winning wasn't particularly random, but how would you argue that in a fair court of law? Traditional probability theory does not even have the notion of random events. Algorithmic information theory does, but it is not applicable to real-world scenarios like the lottery one. We attempt to rectify that.
\end{abstract}

\newpage
\tableofcontents

\newpage
\section{Introduction}
\label{sec:intro}

To motivate our study, we begin with four examples. In each of the four cases, a probabilistic
trial produces a suspicious outcome, and the question arises whether the outcome is the result of
pure chance.  The first case is a thought experiment inspired by a remark of Leonid Levin in
article \cite{Levin:1984}.  The second and third cases are also thought experiments, arguably
more realistic.  The fourth case is real.

\subsection{A Lottery Case}
\label{sub:lottery1}

John and Donna live in Georgia, a state of about 10,000,000 inhabitants.  John is Donna's husband
and the president of the Georgia State Lottery. Anybody may enter into the lottery by buying as
many tickets as (s)he wishes. Every ticket is uniquely numbered.  A winner is chosen at random by
the selection of a ticket number.  This year Donna happens to win.  Over 10,000,000 tickets were
purchased in total, spread among about 4,000,000 people.  Donna purchased three. John bought
none.

Soon after this win is announced, the local media begins to echo claims of corruption against
John. How could it be that of all of the about 4,000,000 participants, the president's wife won?
Surely something must be amiss. John, faced with allegations of unfairness, argues as follows:

\begin{quote}
Someone had to win the lottery.
The process of choosing the winner was fair.
Almost every ticket owner (the only exception being a handful of people who bought many tickets)
had a small chance of winning. If a stranger to me who also bought a small number of tickets had
won, no one would be crying foul. But, such a stranger would have roughly the same small
probability of winning as Donna did. Given that someone had to win, nothing strange has happened.
In particular, there are no grounds to claim the lottery was rigged.
\end{quote}

\noindent
Would you believe him?

\subsection{A Jury Case}
\label{sub:jury1}

Thomas seemed to be a common criminal but there was something uncommon about his case. At least
the prosecutor thought so. As prospective jurors were questioned, she realized that some of them
were unusually informed. She investigated. It turned out that seven out of 50 prospective jurors
belonged to a Facebook group of about 100 people that discussed Thomas's case. This was the only
Facebook group that discussed the case.

Prospective jurors had been chosen at random from a population of about one million adults
available for the purpose. Can it be mere chance that so many prospective jurors belong to the
one and relatively small Facebook group that have discussed the case?

\subsection{A Stalking Case}
\label{sub:stalking1}

Alice and Bob are a couple living in New York City. They don't have a real kitchen and usually
dine out.  Chris is Alice's unstable ex-boyfriend, and  Alice believes that Chris is stalking
her. Too often she has seen him at restaurants.  Alice has wanted to obtain a restraining order
but Bob has argued that they didn't have enough evidence to convince the authorities. After all,
Chris and Alice used to live together and may naturally frequent the same restaurants.  So Bob
suggested an experiment: ``There are at least 100 reasonable restaurants within walking distance
from our place. For the next ten nights, let's pick a restaurant at random except that it should
be a new restaurant each time''.  They performed the proposed experiment. In 5 out of the 10
cases, Chris showed up. Is this evidence sufficient for Alice to obtain a restraining order?

\subsection{The Case of the Man with the Golden Arm}
\label{sub:caputo1}

The story appeared in New York Times on July 23, 1985, on page B1. We learned of it from the book
\cite{Dembski:1998}.

\begin{quote}
TRENTON, July 22 --- The New Jersey Supreme Court today caught up with the ``man with the golden
arm,'' Nicholas Caputo, the Essex County Clerk and a Democrat who has conducted drawings for
decades that have given Democrats the top ballot line in the county 40 times out of 41 times.
\end{quote}

\noindent
The court felt that something was wrong.

\begin{quote}
 The court suggested --- but did not order --- changes in the way Mr. Caputo conducts the
drawings to stem ``further loss of public confidence in the integrity of the electoral
process.''
\end{quote}

\noindent
Caputo wasn't punished. A question arises whether the circumstantial evidence was sufficient to
justify punishing him.

\subsection{Overview}
\label{sub:overview}

The four cases above have something in common. In each case, there is a strong suspicion that a
presumably random event is not random at all. But how can one justify the suspicion?

The purpose of this paper is to build a practical framework in which the desired chance-elimination arguments can be formalized and defended. We start, in \S\ref{sec:bridge}, with a classical principle, often called Cournot's principle, according to which it is a practical certainty that an event with very small probability will not happen. We expound Cournot's principle. In particular, we make explicit that the event of interest is supposed to be specified in advance. Then we generalize Cournot's principle to a more liberal principle, called the bridge principle, that requires only that the event of interest be specified independently from the execution of the probabilistic trial in question. At the end of \S\ref{sec:bridge}, we address the question how an after-the-fact specification can be independent. The inspiration comes from algorithmic information theory, and the intuitive idea is this: some specifications are so succinct that they could have been naturally written ahead of time (and maybe have been written in similar cases in the past).

\S\ref{sec:complexity} is auxiliary. First we recall some basic notions of algorithmic information theory, in particular the notion of Kolmogorov complexity (or information complexity) of events. In algorithmic information theory, events are represented by binary strings. The Kolmogorov complexity of an event is the length of a shortest program for a fixed universal Turing machine that outputs the string presentation of the event. This approach does not work for our purposes. In each probabilistic case in question, we need specifications formulated in terms pertinent to the case and abstracted from irrelevant information. To this end we use logic, and the rest of the section is devoted to logic. We recall and illustrate logic structures and the notion of logical definability. Then we introduce and discuss the notion of the description complexity of events.

In \S\ref{sec:impugn}, we explain how we intend to impugn randomness.
In \S\ref{sec:examples} we illustrate our approach on the cases described above.
There are many discussions with our old friend Quisani throughout the article. The final discussion is in \S\ref{sec:final}.

\subsection*{Acknowledgments}

We thank Andreas Blass, Leonid Levin, Glenn Shafer and Vladimir Vovk for
discussions that influenced our thinking, and we thank Cristian Calude, Guido de Caso, Alexander Shen and Paul Vit\'{a}nyi for very useful comments on a draft of this paper.

\section{Bridging Probabilities and the Physical World}
\label{sec:bridge}

We explicate and broaden the well known principle according to which it is a practical certainty
that an event with very small probability will not happen.

\subsection{Cournot's Principle}
\label{sub:cournot}

Probability theory is applied widely and successfully. But what makes
this mathematical theory relevant to the physical world?  In The Art of
Conjecturing, published posthumously in 1713
\cite{Bernoulli:1713}, Jakob Bernoulli related
mathematical probabilities to practical (or ``moral'') certainty:

\begin{quote}
  Something is morally certain if its probability is so close to certainty that the shortfall is
imperceptible. Something is morally impossible if its probability is no more than the amount by
which moral certainty falls short of complete certainty. Because it is only rarely possible to
obtain full certainty, necessity and custom demand that what is merely morally certain be taken
as certain. It would therefore be useful if fixed limits were set for moral certainty by the
authority of the magistracy --- if it were determined, that is to say, whether 99/100 certainty
is sufficient or 999/1000 is required.
\end{quote}

\noindent
In other words, it is a practical certainty that an event with very small probability will not
happen.
Antoine Cournot seems to be the first to suggest, in the book \cite{Cournot:1843}, that this
principle is the only way of connecting mathematical probabilities to the world. Accordingly the
principle is often ascribed to Cournot.

\medskip\noindent
{\bf Cournot's Principle}\quad
It is a practical certainty that an event with very small probability will not happen.
\medskip

The principle was supported by many heavyweights of probability theory and statistics, in particular \'{E}mile Borel \cite{Borel:1943}, Ronald A. Fisher \cite{Fisher:1925}, Andrei N. Kolmogorov \cite{Kolmogorov:1933} and Paul L\'{e}vy \cite{Levy:1925}. Borel called Cournot's principle The Single Law of Chance.

More information on Cournot's principle is found in the Shafer and Vovk book
\cite{Shafer-Vovk:2001} and Shafer's lecture \cite{Shafer:2006} that is available online.

\subsection{How Small is Sufficiently Small?}
\label{sub:small}

\smallskip\noindent\textbf{Quisani:\ } How small is sufficiently small? I presume that there is some cut-off point, a threshold value, the least upper bound for the sufficiently small values.

\smallskip\noindent\textbf{Authors:\ } Yes, that's the idea. Let us quote \'{E}mile Borel in this connection. In his 1943 book \cite{Borel:1943} for the non-scientist he wrote the following.

\begin{quote}
When we stated The Single Law of Chance, ``events whose probability is sufficiently small never occur,'' we did not conceal the lack of precision of the statement.  There are cases where no doubt is possible; such is that of the complete works of Goethe being reproduced by a typist who does not know German and is typing at random.  Between this somewhat extreme case and ones in which the probabilities are very small but nevertheless such that the occurrence of the corresponding event is not incredible, there are many intermediate cases.  We shall attempt to determine as precisely as possible which values of probability must be regarded as negligible under certain circumstances.
\end{quote}

\noindent
Note ``under certain circumstances''. Different application areas may use different threshold values.  You may want to reject the null hypothesis ``by preponderance of evidence'' or ``beyond a reasonable doubt''. If the courts of law were to use probabilistic thresholds (tailored to specialized circumstances), these distinct judicial criteria would give rise to distinct threshold values. Different criteria and different threshold values may be used in testing scientific hypotheses.

\Q I suppose much experience is needed to just propose a reasonable threshold for a fixed application area. And new developments, new technologies may require that the accepted value be revised. I am thinking of the use of DNA evidence in courts. A number of convictions have been overturned. If the courts used thresholds, some of them should have been adjusted down. I can imagine also the necessity to adjust a threshold value up. Think of cases when people have not been convicted but later confessed to crimes.

\A The issue of appropriate threshold values has been much discussed, especially in connection with statistical hypothesis testing. In fact there are many discussions, in various applications domains, e.g. clinical trials \cite{Davis-Hardy:1990,Friedman:1996,Meinert:1986,Moye:1998}, psychology \cite{Cohen:1994,Fraley:2003}. Statistical hypothesis testing is often misunderstood and abused \cite{Batanero:2000,Cohen:1994,Fraley:2003,Morrison+1:1970}. In the rest of the paper, we will avoid the issue of appropriate threshold values.

\begin{proviso}
Given a probabilistic trial, we will always assume the existence of an agreed and current probability threshold for the application domain of the trial.
\end{proviso}

\subsection{Cournot's Principle Expounded}
\label{sub:expound}

Our formulation of Cournot's principle is rather common. It is also aphoristic. As stated, Cournot's principle does not hold water: events of very small probability happen all the time in the physical world. Several aspects of the principle are implicit. One aspect, for example, is that the probabilistic experiment in question is performed only once. (This particular aspect is made explicit in \cite{Kolmogorov:1933}.) In this subsection we explicate Cournot's principle. Later we will broaden it and call it the bridge principle to emphasize that it bridges between probability theory and the physical world.

As in probability theory, a trial $T$ is a real or imaginary experiment with a well defined set $\Omega_T$ of possible outcomes and with events of $T$ as subsets of $\Omega_T$. In general (when $\Omega_T$ is uncountable) there may be some subsets of $\Omega_T$ that are not events. The case of most interest to us is when $\Omega_T$ is finite; in that case every subset of $\Omega_T$ is an event.

We introduce a few nonstandard terms that are useful for our purposes. An \emph{executed trial} is a trial together with a particular execution of the trial; the execution results in a particular outcome called the \emph{actual outcome} of the executed trial. An event $E$ \emph{happens} or \emph{occurs} at  the executed trial if and only if $E$ contains the actual outcome.

A \emph{probabilistic trial} $(T,\F)$ is a trial $T$ together with a hypothesis, called the \emph{null hypothesis}, that the probability distribution that governs the trial $T$ belongs to $\F$. The probability distributions of $\F$ are the \emph{innate probability distributions} of the probabilistic trial $(T,\F)$. (We are not going to define what it means for a trial $T$ to be governed by a probability distribution $\P$; the connection to the physical world is given by the expounded Cournot principle below.) An \emph{executed probabilistic trial} is a probabilistic trial $(T,\F)$ together with a particular execution of $T$ (that produces the actual outcome of the executed probabilistic trial).

An event $E$ of a probabilistic trial $(T,\F)$ is \emph{negligible} if, for every innate probability distribution $\P$, the probability $\P(E)$ is less than the current probability threshold in the application area of the trial.

\begin{example}\label{ex:1}
View the lottery of \S\ref{sub:lottery1} as an executed probabilistic trial with possible outcomes of the form ``$o$ wins the lottery'' where $o$ ranges over the lottery ticket owners. The null hypothesis says that the trial is governed by the probability distribution where the probability of outcome ``$o$ wins the lottery'' is proportional to the number of lottery tickets that $o$ owns. The actual outcome is ``Donna wins the lottery''. Since Donna bought three tickets, the event ``the winner bought three tickets'' occurs during the execution.
\end{example}

A \emph{probabilistic scenario} $(T,\F,E)$ is a probabilistic trial $(T,\F)$ together with an event $E\subseteq\Omega_T$ called the \emph{focal event} of the probabilistic scenario. An \emph{executed probabilistic scenario} is a probabilistic scenario $(T,\F,E)$ together with a particular execution of $T$ (that produces the actual outcome of the executed probabilistic scenario).

\paragraph{Cournot's Principle Expounded}
Consider a probabilistic scenario with a negligible focal event. If the focal event is specified
before the execution of the trial then it is practically certain that the focal event will not
happen upon the execution.
\medskip

\Q What is the point in fixing an event and execution?

\A If too many small-probability events are specified then it may become likely that at least one
of them happens even if we restrict attention to one execution of the trial. Similarly any event
of  positive probability becomes likely to happen if the trial is  executed too many times.

\Q Given an informal description of a trial, it may be not obvious what the possible outcomes
are. Consider the lottery case. The way the story is told in \S\ref{sub:lottery1}, every outcome
is associated with the winning person. This is natural. But it is also natural, maybe even more
natural, to associate outcomes with the winning tickets.

\A You are right. An informal description of a trial may be somewhat vague about precisely what possible outcomes are. But the definition of a probabilistic trial requires that the set of possible outcomes be indicated explicitly. In the lottery case, there are indeed these two natural ways to view possible outcomes. It does not really matter which way to go. We picked the first way because it is a tiny bit more convenient for our purposes.

\subsection{Cournot's Principle and Statistical Hypothesis Testing}
\label{sub:fisher}

\Q You started with the problem of connecting probabilities to the physical world. But do you
know the true probabilities of real world events? I think not. All you have is a mathematical
model. It surely is at best an approximation of reality. For example, people speak about tossing
a fair coin but no real coin is perfectly fair and no tossing is perfect. More importantly, a
real world trial may be rigged, so that your mathematical model may be far from reality. I would
not be surprised if some magicians can produce any desired sequence of heads and tails by
repeatedly tossing a coin.

\A Turn Cournot's principle upside down. Consider an executed probabilistic scenario with a
negligible focal event specified before executing the trial. If the focal event occurs during the
execution of the trial then reject the null hypothesis.

\Q Is this related to statistical hypothesis testing?

\A This is, as far as we understand, the basic idea of Ronald A. Fisher's method of statistical hypothesis testing  \cite{Fisher:1925,Fisher:1926,Fisher:1956}. The term ``null hypothesis'' is borrowed from Fisher. In Fisher's approach, the focal event is specified by means of the $p$-value of some statistics.

\Q Hmm, I don't know anything about $p$-values. What is your best reference on Fisher's
approach?

\A The Cox and Hinkley book \cite{Cox-Hinkley:1974}.

\subsection{The Bridge Principle}
\label{sub:bridge1}

Now we are ready to broaden Cournot's principle. We use the definitions of \S\ref{sub:expound}.

\medskip\noindent
{\bf The Bridge Principle}
Consider a probabilistic scenario with a negligible focal event. If the focal event is specified
\emph{independently} of the  execution of the trial then it is practically certain that the focal
event does not happen upon the execution.
\medskip

We will use the bridge principle as a ground for rejection of (or at least as a significant argument against) the null hypothesis.

\Q How can an after-the-fact specification be independent? I think that I understand the intent of prior specifications.  They are predictions.  If I give Grant a deck of cards and, without examining the cards, he draws the king of spades, there is no surprise. But if he announces in advance that he is going to draw the king of spades and then indeed he does that, then there is a surprise. I'd think that Grant is a magician. But after-the-fact ``predictions'' do not make sense to me.

\A Suppose that Grant did not announce a card in advance.  Instead Yuri announces the card, after the fact and apparently without any  communication with Grant. Is there an element of surprise?

\Q Yes, I suppose there is. I would suspect that Grant is a magician or that there was some communication between you two. I guess the point is not that the focal event is specified in advance but that it is specified --- possibly after the fact --- without any knowledge about the outcome of the trial, not even partial knowledge.

\A Let us give a name to the principle that you propose:

\medskip\noindent
{\bf The Narrow Bridge Principle} Consider a probabilistic scenario with a negligible focal event. If the focal event is specified without any information about the actual outcome of the trial then it is practically certain that the focal event does not happen upon the execution.
\medskip

\noindent
Unfortunately the narrow bridge principle is too narrow for our purposes. To illustrate the broader principle, consider a trial that consists of tossing a fair coin 41 times. Would the outcome

\begin{equation}\label{equ:ht}
\mathrm{HTTT THTH HTHT HTHH TTHT THHH TTTH HTHH TTTT THHH H}
\end{equation}
\noindent
be surprising?

\Q I do not think so.

\A And what about the outcome where all tosses but the last one came up heads?

\Q Yes, it would be surprising.  I would suspect cheating.

\A But why? Both outcomes have exactly the same probability, $2^{-41}$.

\Q I begin to see your point. The second outcome is special.  It is surprising even though it has not been predicted. I guess it is surprising by its very nature.

\A Yes. But what makes the second outcome special?

\Q The particularly simple specification?

\A That is exactly it. The particularly simple specification makes the outcome surprising and independent from the execution of the trial.

\Q But how do you measure the simplicity of specifications? There are 41 characters, including blanks, in the phrase ``all tosses but the last one came up heads'', and 41 binary symbols in \ref{equ:ht}. Since the Latin alphabet (with the blank symbols) is richer than the binary alphabet $\{H,T\}$, one can reasonably argue that the specification \ref{equ:ht} is simpler, much simpler.

\A This is a good question.  The whole next section is devoted to it.

\section{Random Events and their Specification \\ Complexity}
\label{sec:complexity}

\subsection{Algorithmic Information Theory}
\label{sub:ait1}

The introductory examples illustrate the possibility that some presumably random events may be not random at all.

\Q What does it mean that an event is random?

\A Classical probability theory does not address the question but algorithmic information theory (AIT) does. The basic ideas of AIT were discovered in the 1960s independently by Ray Solomonoff \cite{Solomonoff:1964}, Andrei N. Kolmogorov \cite{Kolmogorov:1965} and Gregory J. Chaitin \cite{Chaitin:1966}. The history of these discoveries is described in \S1.13 of the book \cite{Li-Vitanyi:1997} by Li and Vit\'{a}nyi that we use as our main reference on the subject. Chaitin's book  \cite{Chaitin:1987} on AIT is available online.

A key notion of AIT is the Kolmogorov (or information) complexity of strings.  Intuitively, the Kolmogorov complexity $C(s)$ of a string $s$ is the length of a shortest program that outputs $s$. The larger $C(s)$ is, comparative to the length of $s$, the more random $s$ is.

\Q Programs in what programming language?

\A Programs for a universal Turing machine. AIT was influenced by the computation theory of the time. Traditionally, in AIT, one restricts attention to Turing machines with the binary alphabet, and the universality of a Turing machine $U$ means that $U$ faithfully simulates any other Turing machine $T$ on any input $x$ given $T$ (in one form or another) and given exactly that same input $x$. View a universal Turing machine $U$ as a programming language, so that programs are binary strings.  The Kolmogorov complexity $C_U(s)$ of a binary string $s$ is the length of a shortest $U$ program that outputs $s$.

\Q But this depends on the choice of a universal Turing machine $U$.

\A It does. But, by the Invariance Theorem \cite[\S2.1]{Li-Vitanyi:1997}, for any two universal Turing machines $U_1$ and $U_2$, there is a constant $k$ such that
$
 C_U(s) \leq C_V(s) + k
$
for all binary strings $s$.  In that sense, the dependence on the choice of universal Turing machine $U$ is limited.

There is also a conditional version $C_U(s|t)$ of Kolmogorov complexity, that is the complexity of string $s$ given a string $t$.

\Q I wonder how would one use Kolmogorov complexity to show that the suspicious outcomes of the introductory examples are not random.

\A Unfortunately Kolmogorov complexity does not seem to work well for our purposes.
Whatever universal Turing machine $U$ is fixed, the function $C_U(s)$ is not computable \cite[Theorem~2.3.2]{Li-Vitanyi:1997}. And the machine $U$ does not know anything about the scenarios. Consider the lottery scenario for example. Intuitively the event of Donna winning the lottery should have smaller description complexity than the event of some stranger to John winning the lottery. But this is most probably not the case, precisely because the machine $U$ does not know anything about the scenario.

\Q Maybe one can use the conditional version $C_U(s|t_0)$ of Kolmogorov complexity where $t_0$ is a particular string that describes the given scenario. I suspect that the function $f(s) = C_U(s|t_0)$ is still uncomputable. But maybe one can approximate it.

\A Maybe. But it seems to us that there are simpler and more natural ways to deal with scenarios like those in our introductory examples.

\subsection{Description Complexity}
\label{sub:dc}

We assume that the reader is familiar with the basics of first-order logic though we recall some
notions. A (one-sorted) relational structure $A$ consists of
\begin{itemize}
\item a nonempty set, the \emph{base set} of structure $A$, whose elements are called the
elements of $A$;
\item several relations over the base set including equality, the \emph{basic relations} of
structure $A$; each basic relation has its own arity (the number of arguments);
\item several distinguished elements of $A$ known as \emph{constants}.
\end{itemize}
The ever present equality is a \emph{logic relation}, in contrast to other basic relations. The names of the basic non-logic relations and constants form the \emph{vocabulary} of $A$. Equality is typically omitted when a structure is described. For example, a directed graph is typically described as a relational structure with one binary relation and no constants.

A multi-sorted relational structure is defined similarly except that the base set is split into several nonempty subsets called \emph{sorts}. Each argument position $i$ of every basic relation $R$ is assigned a particular sort $S_i$; the \emph{type} of $R$ is the direct product $S_1\times\cdots\times S_r$ where $r$ is the arity of $R$. Equality, the only logic relation, is an exception. Its type can be described as $\bigcup_S (S\times S)$ where $S$ ranges over the sorts. The vocabulary of a multi-sorted relational structure contains the names of sorts,  relations and constants. Besides, the vocabulary indicates the types of basic relations, individual variables and constants.

\begin{example}\label{ex:2}
Here is a structure related to the lottery scenario. It has two sorts. One sort, called Person, consists of people, namely all lottery ticket owners as well as John, the lottery organizer. The other sort, called Ticket, consists of all the lottery tickets that have been sold. The structure has a binary relation Owns of type Person $\times$ Ticket, with the obvious interpretation. It also has a constant John that denotes the lottery organizer.
\end{example}

If $A$ is a relational structure, $S$ is a sort of $A$ and $X\subseteq S$, we say that $X$ is \emph{definable} in $A$ if there is a first order formula $\phi(x)$ with a single free variable $x$ of type $S$ such that $X$ is the set of elements $a$ of sort $S$
satisfying the proposition $\phi(a)$ in $A$, that is if
$$
 X = \{a:\ A\models\phi(a)\}.
$$
The formula $\phi(x)$ is a \emph{definition} of $X$ in $A$.
The description complexity of $X$ in $A$ is the length of the shortest
definition of $X$ in $A$.

\begin{example}\label{ex:3}
The event ``the winner owns just one ticket'' consists of the outcomes ``p wins the lottery'' where $p$ ranges over the people owning exactly one
ticket. The event is thus definable by formula
$$
 \exists t_1(\Owns(p,t_1) \land (\forall t_2 (\Owns(p,t_2) \to t_1 = t_2))).
$$
in the structure of Example~\ref{ex:2}.
\end{example}

\Q How do you measure the length of a formula?

\A View the names of relations, variables and constants as single symbols, and count the number of symbols in the formula. Recall that the vocabulary specifies the type of every variable and every constant.

\Q As far as the lottery case is concerned, the structure of Example~\ref{ex:2} is poor. For example, it does not distinguish between people that own the same amount of tickets. In particular, it does not distinguish between Donna and anybody who owns exactly three tickets. You can extend it with a constant for Donna. If you want that the structure reflects a broader suspicion that John may cheat, you can add a constant for every person such that there is a reasonable suspicion that John will make him a winner. Much depends of course on what is known about John. For example, you can add a constant for every close relative and every close friend of John.

\A Alteratively we may introduce binary relations CloseRelative$(p,q)$ and CloseFriend$(p,q)$.

\begin{example}\label{ex:4}
Extend the structure of Example~\ref{ex:2} with binary relations
CloseRelative$(p,q)$ and CloseFriend$(p,q)$ of type
Person $\times$ Person, with the obvious interpretations: $q$ is a close relative of $p$, and $q$
is a close friend of $p$ respectively; in either case $q$ is distinct from $p$.
\end{example}

\subsection{Alternatives}
\label{sub:alt}

\Q There is something simplistic about Example~\ref{ex:4}. Both relations seem to play equal roles. In reality, one of them may be more important. For example, John may be more willing to make a close relative, rather than a close friend, to win. People often put different weights on different relations. For a recent example see \cite{Richardson-Domingos:2006}. You should do the same.

\A You are right of course but, for the time being, we keep things simple.

\Q And why do you use relational first-order logic? There are many logics in the literature.

\A In this first paper on the issue, it is beneficial for us to use relational first-order logic as our specification logic. It the best known and most popular logic, and it works reasonably well.  As (and if) the subject develops, it may be discovered that the best specification logic for one application domain may not be the best for another. At this point, our experience is very limited.

\Q First-order logic isn't the best logic for all purposes.

\A It is not. And there are two distinct issues here. One issue is expressivity. If, for example, you need recursion, first-order logic may be not for you. It lacks recursion. The other issue is succinctness. It is possible to increase the succinctness of relational first-order specifications without increasing the expressive power of the logic. For example, one may want to use function symbols. One very modest extension of relational first-order logic which is nevertheless useful in making specifications shorter is to introduce quantifiers $\dot\exists x\phi$ (note the dot over $\exists$) saying that there exists $x$ different from all free variables of the formula $\phi$ under quantification. If $y_1,\dots,y_k$ are the free variables of $\phi$  then $\dot{\exists}x\phi$ is equivalent to
$$
 \exists x (x\neq y_1 \land \cdots \land x\neq y_k \land \phi)
$$
but it is shorter. It would be natural of course to introduce the $\dot{\exists}$ quantifier together with its dual $\dot{\forall}$ quantifier.

\Q Instead of logic, one can use computation models, especially restricted computation models, e.g. finite state automata, for specification.

\A Yes, you are right. Note though that, for every common computation model, there is a logic with equivalent expressivity. For example, in the case of finite automata over strings, it is existential second-order logic \cite{Buchi:1960}.

\section{Impugning Randomness}
\label{sec:impugn}

Now we are ready to explain our method of impugning the null hypothesis in executed probabilistic scenarios with suspicious outcomes. Given
\begin{itemize}
\item a trial such that some of its outcomes arouse suspicion  and
\item a null hypothesis about the probability distribution that governs the trial,
\end{itemize}
one has several tasks to do.

\medskip\noindent
{\bf 1: Background Information}\quad
Analyze the probabilistic trial and establish what background information is relevant.

\medskip\noindent
{\bf 2: Logic Model}\quad
Model the trial and relevant background information as a logic structure.

\medskip\noindent
{\bf 3: Focal Event}
Propose a focal event that is
\begin{itemize}
\item negligible under the null hypothesis and
\item has a short description in the logic model.
\end{itemize}

\medskip\noindent
By the bridge principle, the focal event is not supposed to happen, under the null hypothesis, during the execution of the trial. If the focal event contains the actual outcome of the trial, then the focal event has happened. This gives us a reason to reject the null hypothesis.

\subsubsection*{What background information is relevant?}

Relevant background information reflects various ways that suspicious outcomes occur. In this
connection historical data is important. In the lottery case, for example, it is relevant that
some lottery organizers have been known to cheat.

\subsubsection*{What does the Model Builder Know about the Actual Outcome?}

The less the model builder knows about the actual outcome the better. Ideally the model builder has no information about the actual outcome, so that we can use the narrow bridge principle. We may not have a model builder with no information about the actual outcome; it may even happen that the actual outcome has been advertised so widely that everybody knows it. In the absence of blissfully unaware model builder, we should try to put ourselves into his/her shoes.

\subsubsection*{The Desired Logic Model}

One may be lucky to find an existing logic model that has been successfully used in similar
scenarios. If not, construct the most natural and frugal model you can.

\Q ``Natural'' is a positive word. Surely it is beneficial that the desired model is natural.
But why should the model be frugal?

\A If the model is too rich (like in the case of classical algorithmic information theory), too
many events have short specifications. Imagine for example that, in the lottery case, the model
allows you to specify shortly various people that have nothing to do with the lottery organizer.

\Q But how do you know that those people have nothing to do with the lottery organizer? Maybe one
of them is a secret lover of the lottery organizer.

\A Indeed, the background information deemed relevant may be deficient. But, at the model
building stage, we want to reflect only the background information deemed relevant.

\subsubsection*{The Desired Focal Event}

The desired focal event contains the suspicious outcomes of the trial.

\section{Examples}
\label{sec:examples}

We return to the four cases of \S\ref{sec:intro}.

\subsection{The Case of Lottery}
\label{sub:lottery2}

\paragraph{Trial} For an informal description of the trial see \S\ref{sub:lottery1}. Recall that
John is the lottery organizer, and Donna is his wife. As mentioned in Example~\ref{ex:1}, we view
the lottery as a trial with potential outcomes of the form ``o wins the lottery'' where o ranges
over the lottery ticket owners.

\paragraph{Null Hypothesis}

There is only one innate probability distribution $\P$, and the probability $\P(x)$ of any person
$x$ to win is proportional to the number of lottery tickets that $x$ owns.

\paragraph{Background Information} We assume that the following is known about John, the lottery organizer. He is a family man, with a few close friends that he has known for a long time. He bought no lottery tickets.

\paragraph{Actual Outcome} ``Donna wins the lottery''.

\paragraph{Logic Model} Our model is a simplification of the structure of Example~\ref{ex:4}. We don't need the sort Ticket introduced originally in Example~\ref{ex:2} to illustrate the notion of multi-sorted model. And we do not need the full extent of relations CloseRelative and CloseFriend, only the sections of them related to John. Our model is one-sorted. The one sort, called Person, consists of people, namely all lottery ticket owners as well as John, the lottery organizer. The structure has one constant and two unary relations. The constant is John; it denotes the lottery organizer. The two unary relations are $\CloseRelativeOfJohn(p)$ and $\CloseFriendOfJohn(p)$, both of type Person. The interpretations of the two relations are obvious: $p$ is a close relative of John, and $p$ is a close friend of respectively; in both cases $p$ is distinct from John.

\paragraph{Focal Event:} The winner is a close relative or friend of John, the lottery organizer, in other words, the winner belongs to the set
$$
 \{x:\ \CloseRelativeOfJohn(x) \lor \CloseFriendOfJohn(x) \}.
$$

\subsection{The Case of Jury Selection}
\label{sub:jury2}

\paragraph{Trial}

For an informal description of the trial see \S\ref{sub:jury1}. The trial in question selects a
pool of 50 prospective jurors from about a 1,000,000 people available for the purpose.

\paragraph{Null Hypothesis}

There is only one innate probability distribution, and the one innate probability distribution is
uniform so that all possible pools of 50 prospective jurors are equally probable.

\paragraph{Background Information}

There is a unique Facebook group of about 100 people that discusses the criminal case.

\paragraph{Actual Outcome} A pool with seven prospective jurors from the Facebook group.

\paragraph{Logic Model}

The model has two sorts and one relation.
\begin{itemize}
\item Sort Pool consists of all possible pools of 50 prospective jurors.
\item Sort Member consists of the members of the Facebook group that discussed Thomas's
case.
\item The relation $\In(m,p)$ of type Member $\times$ Pool holds if and only if member $m$ of the Facebook group belongs to
pool $p$.
\end{itemize}

\paragraph{Focal Event}

$$
 \{p:\ \exists m_1 \exists m_2
 \big(m_1\neq m_2\ \land\ \In(m_1,p)\ \land\ \In(m_2,p) \big)\}
$$

\Q If the null hypothesis is impugned then some rules have been violated. Who is the guilty
party? In the lottery case, it was clear more or less that John was the guilty party. In this
case, it is not obvious who the guilty party is.

\A We do not pretend to know the guilty party. Our only goal is to impugn randomness. There may
be more than one guilty party as far as we know. Their actions may or may not have been
coordinated.

\Q As far as I see, plenty of randomness might have remained. You did not impugn all randomness.

\A You are right. Let us express our goal more precisely: it is to impugn the null hypothesis, no
more no less.

\subsection{A Stalking Case}
\label{sub:stalking2}

\paragraph{Trial}
An informal description of the case is given in \S\ref{sub:stalking1}. The 10 nights of the trial
may be represented by numbers $1,\dots,10$.
The outcomes of the trial may be represented by functions $f$ from $\{1, \ldots, 10\}$ to
$\{0,1\}$ where the meaning of $f(n)=1$ (resp. $f(n)=0$) is that Alice and Bob met (resp. did not
met) Chris at the restaurant on night $n$.

\paragraph{Null Hypothesis}

Intentionally, the null hypothesis says that Chris does not stalk Alice. Formally, the null
hypothesis says that a probability distribution $\P$ on the outcomes is innate if and only if it
satisfies the following two requirements for every outcome $f$.
\begin{enumerate}
\item Events $f(n_1)=1$ and $f(n_2)=1$ are independent for any nights $n_1\neq n_2$.
\item $\P(f(n)=1) \leq 1/100$ for every night $n$.
\end{enumerate}
\noindent
Requirement 2 says that $\P(f(n)=1)$ is less than (rather than equal to) $1/100$ rather than
$\P(f(n)=1) = 1/100$ because, at night $n$, Chris may not to be present at all in any of the 100
restaurants at the time when Alice and Bob dine. If he is in one of the 100 restaurants when
Alice and Bob dine then $\P(f(n)=1) = 1/100$.

\paragraph{Background Information} Chris is suspected of stalking Alice in restaurants.

\paragraph{Actual Outcome} Five times out of ten times Alice and Bob meet Chris at the chosen
restaurant.

\paragraph{Logic Model}
There are three sorts of elements and one relation.
\begin{itemize}
 \item Sort Night consists of numbers $1,\dots,10$.
 \item Sort Outcome consists of all possible functions from Night to $\{0,1\}$.
 \item Relation $R(f,n)$ holds if and only if, on night $n$, Alice and Bob meet Chris at the
chosen restaurant.
\end{itemize}

\noindent
{\bf Focal Event}\quad
 $\{f:\ \exists n_1 \exists n_2
 \big( n_1\neq n_2 \land R(f,n_1) \land R(f,n_2) \big) \}$.
\medskip

\Q The focal event is that there are two distinct nights when Chris dines at the restaurant where
Alice and Bob dine. What if the current probability threshold is lower than you presume, and the
focal event turns out to be non-negligible?

\A In this particular case, it is natural to consider the focal event that there are three
distinct nights when Chris dines at the restaurant where Alice and Bob dine.

\Q It is rather expensive to say that there are $k$ distinct elements; the description complexity is $O(k^2)$. Now I see why you mentioned those dotted existential quantifiers $\dot{\exists}$ in \S\ref{sub:alt}.

\subsection{The case of Nicholas Caputo}
\label{sub:caputo2}

\paragraph{Trial} An informal description of the case is given in \S\ref{sub:caputo1}. The $41$ elections may be represented by numbers $1,\dots,41$. The possible outcomes of the trial can be seen as functions $f$ from $\{1,\dots,41\}$ to \{D,N\} where the letters $D$, $N$ indicate whether the top ballot line went to a Democrat or not.

\paragraph{Null Hypothesis}
Intentionally the null hypothesis is that the drawings were fair.
Formally, the null hypothesis says that there is a unique innate probability distribution $\P$ on
the outcomes, and that $\P$ satisfies the following two requirements for every outcome $f$.
\begin{enumerate}
\item Events $f(e_1)=D$ and $f(e_2)=D$ are independent for any elections $e_1\neq e_2$.
\item For every election $e$, $\P(f(e)=D) = r_e$ where $r_e$ is the fraction of Democrats on
the ballot.
\end{enumerate}
\noindent
We assume that every $r_e\geq 0.4$.

\paragraph{Background Information} The county clerk, who conducted the drawings, was a Democrat.

\paragraph{Actual Outcome} $40$ times out of $41$ times the top ballot line went to Democrats.

\paragraph{Logic Model} The model has three sorts, two constants and one relation.
\begin{itemize}
\item Sort Election consists of numbers $1,\ldots,41$ representing the $41$ elections.
\item Sort Party consists of two elements. The elements of Party will be called parties.
\item Sort Outcome consists of all $f$ from Election to Party.
\item The constants $D$ and $N$ of type Party denote distinct elements of type Party.
\item The relation $R(f,e,p)$ of type Outcome $\times$ Election $\times$ Party holds if and only if, according to outcome $f$, the top ballot line went to party $p$ at elections $e$.
\end{itemize}

\paragraph{Focal Event:}
$$
 \{f:\ \forall i\forall j ((R(f,i,N) \land R(f,j,N))\to i=j)\}.
$$

\section{Related Work}
\label{sec:related}

Our paper touches upon diverse areas of science. We restrict attention to a few key issues: Cournot's principle, algebraic information complexity, and social network analysis.

\subsection{Cournot's Principle}
\label{sub:bridge2}

The idea that specified events of small probability do not happen seems to be fundamental to our human experience. And it has been much discussed, applied and misapplied. We don't --- and couldn't --- survey here the ocean of related literature. In \S\ref{sec:bridge} we gave already quite a number of references in support of Cournot's principle. On the topic of misapplication of Cournot's principle, let us now turn to the work of William Dembski. Dembski is an intelligent design theorist who has written at least two books, that are influential in creationist circles, on applications of ``The Law of Small Probability'' to proving intelligent design \cite{Dembski:1998,Dembski:2006}.

We single out Dembski because it is the only approach that we know which is, at least on the surface, similar to ours. Both approaches generalize Cournot's principle and speak of independent specifications. And both approaches use the information complexity of an event as a basis to argue that it was implicitly specified. We discovered Dembski's books rather late, when this paper was in an advanced stage, and our first impression, mostly from the introductory part of book \cite{Dembski:1998}, was that he ate our lunch so to speak. But then we realized how different the two approaches really were. And then we found good mathematical examinations of the fundamental flaws of Dembski's work: \cite{Wein:2002} and \cite{Bradley:2009}.

Our approach is much more narrow. In each of our scenarios, there is a particular trial $T$ with well defined set $\Omega_T$ of possible outcomes, a fixed family $\F$ of probability distributions --- the innate probability distributions --- on $\Omega_T$, and a particular event --- the focal event --- of sufficiently  small probability with respect to every innate probability distribution. The null conjecture is that the trial is governed by one of the innate probability distributions. Here events are subsets of $\Omega_T$, the trial is supposed to be executed only once, and the focal event is supposed to be specified independently from the actual outcome. By impugning randomness we mean impugning the null hypothesis.

Dembski's introductory examples look similar. In fact we borrowed one of his examples, about ``the man with a golden arm''. But Dembski applies his theory to vastly broader scenarios where an event may be e.g. the emergence of life. And he wants to impugn any chance whatsoever. That seems hopeless to us.

Consider the emergence of life case for example. What would the probabilistic trial be in that case? If one takes the creationist point of view then there is no probabilistic trial. Let's take the mainstream scientific point of view, the one that Dembski intends to impugn. It is not clear at all what the trial is, when it starts and when it is finished, what the possible outcomes are, and what probability distributions need to be rejected.

The most liberal part of our approach is the definition of independent specification. But even in
that aspect, our approach is super narrow comparative to Dembski's.

There are other issues with Dembski's work; see \cite{Wein:2002,
  Bradley:2009}.

\subsection{Algorithmic Information Theory}
\label{sub:ait2}

The idea of basing the intrinsic randomness of an event upon its description in a fixed language is fundamental to algorithmic information theory (in short AIT) \cite{Chaitin:1987,Li-Vitanyi:1997} originated by Ray Solomonoff \cite{Solomonoff:1964}, Andrei N. Kolmogorov \cite{Kolmogorov:1965} and Gregory J. Chaitin \cite{Chaitin:1966}.

In \S\ref{sub:ait1}, we sketched the basic ideas of the theory. In the classical AIT, the theoretical power is gained by basing the information complexity measure on universal Turing machines. This becomes an impediment to practical applications; the classical information complexity of (the string representation of) events is not computable. For practical applications, it is thus natural to look at restricted variants of AIT which ``impoverish'' the event description language even though the classical theorems of AIT may no longer hold.

The influential Lempel-Ziv compression theory of strings \cite{Lempel-Ziv:1977,Lempel-Ziv:1978} can be viewed as such a restriction of AIT. However Lempel and Ziv developed their theory without any direct connection with AIT. One recent and even more restrictive theory \cite{Calude+2:2011} was inspired by AIT: ``we develop a version of Algorithmic Information Theory (AIT)
by replacing Turing machines with finite transducers''.

One useful application of AIT to real-world phenomena has been through the Universal Similarity Metric and its uses in genetics and bioinformatics \cite{Krasnogor:2004, Li:2001, Ferragina:2007, Gilbert:2007}, plagiarism detection \cite{ChenLi:2004} and even analysis of music \cite{Cilibrasi:2004}. In \cite{Cilibrasi:2007}, the authors combine a restricted variant of Kolmogorov complexity with results obtained from Google searches to derive a metric for the similarity of the meaning of words and phrases.  In doing so, they are able to automatically distinguish between colors and numbers, perform rudimentary automatic English to Spanish translation, and even distinguish works of art by properties of the artists. In such lines of research, practitioners often replace the Kolmogorov complexity measure with measures based on string-compression algorithms  \cite{Welch:1984,bzip2,ppm} more efficient than the original Lempel-Ziv algorithms.

In cognitive science, the simplicity theory of Chater, Vit\'{a}nyi, Dessalles and Schmidhuber offers an explanation as to why human beings tend to find certain events ``interesting'' \cite{Chater:2008}. The explanation correlates the interest of an event with its simplicity (i.e., the lowness of its Kolmogorov complexity) .

Our logic-based definition of description complexity in \S\ref{sub:dc} fits this mold of restricted algorithmic information theories. We note, however, that the logic approach is rather general and can handle the classical information complexity and its restricted versions and even its more powerful (e.g. hyper-arithmetical) versions.

\subsection{Social Network Analysis}
\label{sub:social}

The idea of modeling real-world scenarios using relational
structures dates back at least to the 1950s
\cite{Merton:1957}.  The primary scientific
developers of this idea were for many years sociologists and social
anthropologists working in the field of social network analysis.
As a field of mathematical sociology, social network analysis has put forth a network-theory
oriented view of social relationships and used it to quantitatively analyze social phenomena.

Even in the days before massive social network data was available, social network analysts obtained fascinating results. For example, in a 1973 paper ``The Strength of Weak Ties'', Mark Granovetter put forth the idea that most jobs in the United States are found through ``weak ties'', that is acquaintances the job seeker knows only slightly. Granovetter obtained his relational data by interviewing only dozens of people, yet his conclusions held up experimentally and are widely influential today in sociology.
With the advent of large-scale digitized repositories of relational social network data such as Facebook (according to a recent estimate, more than 40\% of the US
population have Facebook accounts \cite{Wells:2010}), the applicability of social network analysis techniques grew tremendously.
The relational algebra of social network analysis tends to be simple.  Typically, analysis is done with rudimentary graph theory: members of a population (called actors) are nodes in a graph and the relationships of interest between actors are modeled as edges. Multiple binary relations are combined into composite relations so that core social network analysis calculations are done over a single graph's adjacency matrix
\cite{Wasserman:1994}.

In the case that our models are graphs, there is much machinery of social network analysis which could be of use to us.  For instance, social network analysts have developed robust and scalable methods for determining the central nodes of interest of social networks, based upon things like weighted connectivity.  We can imagine this being useful for impugning randomness. For instance, if one does not know which members of the population should be distinguished and named by constant symbols, the very structure of a social network may force certain nodes upfront.  There are many other techniques from social network analysis (and available high-performance software) which have the potential to be useful for our goals.

\section{Final Discussion}
\label{sec:final}

\Q I have been thinking about algorithmic information theory and its applications, and I also did some reading, e.g. \cite{Vitanyi+2:1997,Shen:2009}. In general your logic approach appeals to me but I have some reservations.

\A Let's start with the positive part. What do you like about the logic approach?

\Q The situation at hand is described directly and rather naturally. I also like that some outcomes and events are not definable at all. Consider for example the lottery-related model in \S\ref{sub:lottery2}. Unless John, the lottery organizer, has a single close relative or a single close friend, no particular outcome is definable in the model. And the model does not distinguish at all between any two persons outside the circle that contains John, his close relatives and his close friends. This simplicity may be naive but it is certainly appealing.

\A Indistinguishability is important. It is rather surprising in a sense that, in the application of probability theory, so often one is able to compute or at least approximate probabilities. Think about it. The number of possible outcomes of a trial may be rather large and may even be infinite. And these are just outcomes; most events contain multiple outcomes. A probabilistic measure on the event space is a function from the events to real numbers between 0 and 1 that satisfies some slight conditions.

\Q Most probability measures are useless I guess. Which of them are useful?

\A Those few that allow us feasible --- though possibly approximate --- computations of the probabilities of interesting events. Typically useful measures heavily exploit the symmetries inherent in the trial and the independence of various parts of the trial.

\Q I think I see the connection to indistinguishability. But let me go to my reservations. It is basically about the annoying freedom in fixing the probability threshold, in choosing the appropriate logic, in figuring out what background information is relevant and what should the focal event be, in constructing the logical model, and in deciding whether a proposed logical specification of the focal event is short enough.

\A The ``annoying freedom'' is inherent in the impugning-randomness problem.

\Q Kolmogorov complexity is objective, due to the Invariance Theorem mentioned in \S\ref{sub:ait1}.

\A It is objective only to a point.  Recall that the Invariance Theorem involves an unspecified additive constant. So Kolmogorov complexity also suffers from the nagging question ``is it short enough''. Besides, one may be interested in the length of a shortest specification of a given string in first-order arithmetic or Zermelo-Fraenkel set theory for example. The resulting specification complexity measures are rather objective. They are undecidable of course, but so is Kolmogorov complexity.

\Q So how do you intend to deal with the annoying freedom?

\A We believe that the annoying-freedom problem cannot be solved by theorists. It can be solved, better and better, by experimentation, trial and error, accumulation of historical records, standardization, etc.

\Q Allow me one other question before we finish. You mentioned in \S\ref{sub:fisher} that, in Fisher's approach, the focal event is specified by means of the $p$-value of some statistics. ``In statistical significance testing'', says Wikipedia \cite{p-value}, ``the $p$-value is the probability of obtaining a test statistic at least as extreme as the one that was actually observed, assuming that the null hypothesis is true''. Note the closure under the at-least-as-extreme values. If a focal event is not specified by means of a $p$-value, is there any kind of closure that the focal event should satisfy?

\A Yes, in our examples, the focal event contained not only the actually observed outcome but also other suspicious outcomes. In fact, the focal-event approach is rather flexible. Consider the lottery scenario for example. The actual outcome --- that Donna won the lottery --- may be judged to be the most suspicious and, from that point of view, the most extreme, so that there are no other outcomes at least as extreme. But the focal event contains additional suspicious outcomes.

\end{document}